\documentclass[12pt,preprint]{aastex}
\usepackage{amsmath}

\def\sun{\hbox{$\odot$}}
\def\R23{\mbox{$\rm R_{23}$}}

\def\kmsmpc{km s$^{-1}$ Mpc$^{-1}$}

\def\msun{M$_{\odot}$}

\def\Hb{\mbox{${\rm H}{\beta}$}}
\def\Ha{\mbox{${\rm H}{\alpha}$}}
\def\OIIIa{\mbox{${\rm [O\,III]\,}{\lambda\,5007}$}}
\def\OIIIb{\mbox{${\rm [O\,III]\,}{\lambda\,4959}$}}
\def\OII{\mbox{${\rm [O\,II]\,}{\lambda\,3727}$}}

\def\NII{\mbox{${\rm [N\,II]\,}{\lambda\,6584}$}}

\shorttitle{The mass-metallicity and FMR relations at $z > 2$}
\shortauthors{Maier, C. et al.}

\begin{document}

\clearpage
\title{The mass-metallicity and fundamental metallicity relations at $z>2$ using VLT\footnotemark[1] and Subaru\footnotemark[2] near-infrared spectroscopy of zCOSMOS  galaxies}

\author{C. Maier\altaffilmark{3,4}}

\email{christian.maier@univie.ac.at}




\author{
\& S.J.~Lilly\altaffilmark{4},
B.~ Ziegler\altaffilmark{3},
T.~Contini\altaffilmark{5,6},
E.~ P\'erez Montero\altaffilmark{7},
Y.~ Peng\altaffilmark{4,8,9},
I.~ Balestra\altaffilmark{10}
}

\footnotetext[1]{Based on observations
  obtained at the European Southern Observatory (ESO) Very Large
  Telescope (VLT), Paranal, Chile; ESO program 84.B-0232A and large program 175.A-0839}
\footnotetext[2]{Based on data collected at the Subaru telescope, which is operated by the National Astronomical Observatory of Japan (Proposal ID: S09B-013)}
\altaffiltext{3}{University of Vienna, Department of Astrophysics, T\"urkenschanzstrasse 17, 1180 Vienna, Austria}
\altaffiltext{4}{Institute of Astronomy, ETH Zurich, Wolfgang-Pauli-Strasse 27, CH-8093 Zurich, Switzerland}
\altaffiltext{5}{Institut de Recherche en Astrophysique et Plan\'etologie, CNRS, 14 avenue \'Edouard Belin, 31400 Toulouse, France}
\altaffiltext{6}{IRAP, Universit\'e de Toulouse, UPS-OMP, 31400 Toulouse, France}
\altaffiltext{7}{Instituto de Astrof\'isica de Andalucia, CSIC, Apartado de Correos 3004, 18080 Granada, Spain}
\altaffiltext{8}{Cavendish Laboratory, University of Cambridge, 19 J. J. Thomson Avenue, Cambridge CB3 0HE, UK}
\altaffiltext{9}{Kavli Institute for Cosmology, University of Cambridge, Madingley Road, Cambridge CB3 0HA, UK}
\altaffiltext{10}{Max-Planck-Institut f\"ur Extraterrestrische Physik, Postfach 1312, Giessenbachstrasse, D-85741 Garching b. M\"unchen, Germany}


\begin{abstract} 

In the local universe, there is good evidence that, at a given stellar mass $M$, the gas-phase metallicity $Z$ is anti-correlated with the star formation rate ($SFR$) of the galaxies. It has also been claimed that the resulting $Z(M,SFR)$ relation is invariant with redshift - the so-called ``Fundamental Metallicity Relation" (FMR).
Given a number of difficulties in determining metallicities, especially at higher redshifts, the form of the $Z(M,SFR)$ relation and whether it is really independent of redshift is still very controversial.
%
To explore this issue at $z>2$, we used VLT-SINFONI and Subaru-MOIRCS near-infrared spectroscopy of 20 zCOSMOS-deep galaxies at $2.1<z<2.5$ to measure the strengths of up to five emission lines: \OII, \Hb, \OIIIa, \Ha, and \NII. This near-infrared spectroscopy enables us 
to derive
 O/H metallicities, and also SFRs from extinction corrected \Ha\, measurements.
%
We find that the mass-metallicity relation (MZR) of these star-forming galaxies at $z \approx 2.3$ is lower than the local SDSS MZR by a factor of three to five, a larger change than found by \citet{erb06} using [NII]/\Ha-based metallicities from stacked spectra. We discuss how the different selections of the samples and  metallicity calibrations used may be responsible for this discrepancy.  The galaxies show direct evidence that the SFR is still a second parameter in the mass-metallicity relation at these redshifts. However, determining whether the $Z(M,SFR)$ relation is invariant with epoch depends on the choice of extrapolation used from local samples, because $z>2$ galaxies of a given mass have much higher SFRs than the local SDSS galaxies. We find that the zCOSMOS galaxies are consistent with a non-evolving FMR if we use the physically-motivated formulation of the $Z(M,SFR)$ relation from \citet{lilly13}, but not if we use the empirical formulation of \citet{mannu10}.
\end{abstract}





\section{Introduction}
\label{intro}

~~~The $2 < z < 3$ redshift is a critical epoch at which both the global
star formation rate (SFR) and the active galactic nucleus (AGN) activity appear to  have peaked \citep[e.g.][]{hopbea06,fan01}, and the
morphological differentiation of galaxies was fully underway \citep[e.g.][]{pap05,wil09}.
For a period of a few billion years, a large fraction (at least 50\%)
of massive galaxies with $10^{10} < M/M_{\odot} < 10^{11}$
formed stars at sufficiently high SFRs ($\sim 20-100  M_{\odot}/yr$) to produce the bulk of their stellar populations in this span of time: i.e.,  this 
is the ''formation epoch'' of these galaxies. 
Therefore, learning as much as possible about the (physical) properties of these
$z>2$ galaxies will help us to determine how galaxies accumulate their mass and stellar populations.

~~~The specific SFR (SSFR) of most star-forming galaxies has been found to be a tight but weak function of stellar mass
at all epochs up to at least $z \sim 2$, forming the so-called \emph{Main Sequence (MS)} of galaxies \citep[e.g.,][]{brinchmann04,peng10}.
The characteristic SSFR of the Main Sequence increases strongly with redshift, by a factor of 20 back to $z\sim2$ \citep{daddi07,elbaz07,noeske07,salim07,whit12}.    This implies that
galaxies with SFRs that are an order of magnitude or more higher than those of local galaxies of the same stellar mass \citep[see Eq. 1
in][]{peng10} are nevertheless the ``normal'' MS population of
star-forming galaxies at $z>2$.  The tightness of the SSFR on the MS suggests that the SFR is evolving in a quasi-equilibrium state \citep[e.g.,][]{bouchet10,lilly13,birrer14,peng14,pipino14,dekman14}.

~~~The \emph{chemical evolution} of such galaxies, as traced by their gaseous O/H ratio, is a key diagnostic. 
In the local universe, there is a tight 
mass-metallicity relation \citep[MZR,][]{lequeux79} 
with, in SDSS, a 1$\sigma$ dispersion of O/H 
ranging from 0.20 dex at low mass, log(M/\msun )$\sim9$, to
0.07 dex at high mass, log(M/\msun )$\sim11$ \citep{trem04}.
Considering their median errors in determining
masses and metallicities of 0.09 and 0.03 dex, respectively,  \citet{trem04} estimated that roughly half of the
spread in the mass-metallicity relation could be attributed to observational uncertainties.
The SDSS MZR has a relatively steep slope below $M\sim 10^{10.7}$\msun, for a \citet{salp55} IMF, but flattens at higher masses.
Using chemical evolution models,  \citet{trem04} interpreted
the flattening in terms of efficient galactic winds that remove
metals from galaxies with shallow potential wells (lower-mass galaxies), an idea first introduced by \citet{lars74}.  Other explanations are also possible. Flattening of the slope of the MZR at lower redshift has
been interpreted as the chemical version of galaxy downsizing \citep[e.g.][]{maier06,maiol08,zahid11}. On the other hand, \citet{zahid13} claimed that the flattening of the MZR at later times could be due to a redshift evolution in the stellar mass at which the metallicities of galaxies begin to saturate.

It is also well-established that the MZR at $z >> 0$ evolves relative to that seen locally \citep[e.g.,][]{savaglio05,erb06,maier06,maiol08}.
However, the observed shape and evolution of the MZR at high redshifts is likely to be affected by the selection of the samples and metallicity estimators used. 


~~~ In the local universe, the O/H at a given mass also depends on the SFR (and thus also SSFR) of the galaxy
\citep[e.g.,][]{ellison08,mannu10,laralop10,yates12,andmar13},
i.e., the SFR appears to be a ``second-parameter'' in the MZR.  The evidence for an inverse correlation between SFR and O/H is good at low masses, and the scatter in O/H can be reduced to about 0.07 dex across the (M,SFR) plane \citep[][Ma10 in the following]{mannu10}.   However, the form of the Z(M,SFR) relation at higher masses is less well-established, and in fact \citet{yates12}
have claimed that the SFR-dependence reverses. 
Ma10 represented the local SDSS Z(M,SFR) by a second-order polynomial in M and SFR (their Eq.\,2 describing a surface in the 3D space), and, 
by introducing the new quantity $\mu_{0.32}=\rm{logM}-0.32 \times \rm{logSFR}$, they also defined a projection of the Z(M,SFR) that minimizes the metallicity scatter of local galaxies.

~~~Furthermore, Ma10 claimed that the $Z(M,SFR)$ relation seen in measurements of low redshift SDSS galaxies (with $\rm{SFRs}<10M_{\odot}/\rm{yr}$) is also applicable to much higher redshift galaxies, in which the typical SFRs are higher because of the evolution in the MS SSFR.  They coined the phrase Fundamental Metallicity Relation (FMR) to denote this epoch-invariant $Z(M,SFR)$ relation, and claimed that, 
``in practice, metallicity of star-forming galaxies of any mass, any SFR and at any redshift 
up to z=2.5 follow the following relation:
\begin{equation}
\label{eq:cfr2}
\begin{array}{rll}
12+log(O/H)=&8.90+0.47x  & ~\rm{if}~~ \mu_{0.32}<10.2 \\
            &9.07        & ~\rm{if}~~ \mu_{0.32}>10.5 \\
\end{array}
\end{equation}
with $x=\mu_{0.32}-10$.''
This Eq.\,5 of Ma10 describes a plane in 3D space, and is a simplification of their Eq.\,4.

In this paper we clearly distinguish between (a) the existence of a $Z(M,SFR)$ relation at a given epoch (i.e., whether the SFR is a second parameter in the MZR) and (b) whether the $Z(M,SFR)$ relation is also epoch-independent.  We reserve the term FMR for an epoch-{\it independent} $Z(M,SFR)$ relation.  As we discuss further below, it is clear that assessing the epoch-(in)dependence of the $Z(M,SFR)$ relation depends on the choice of extrapolation of any locally-defined $Z(M,SFR)$ relation into regions of the (M,SFR) parameter space that are poorly, if at all, populated in the local universe.  Ma10  suspected that their FMR held up to $z \sim 2.5$, but evolved at $z>3$, using the \citet[][Eb06 in the following]{erb06} sample at $z>2$, and the galaxy samples at $z>3$ of  \citet{maiol08} and \citet{mannu09}.

Initial explanations of why SFR is a second parameter in the MZR have generally involved ad hoc inflow of gas diluting the metallicity and increasing star-formation \citep{mannu10,dave12,dayal13}.
In a recent paper, however, \citet[][Li13 in the following]{lilly13} showed that a $Z(M,SFR)$ relation is a natural outcome of a simple model of galaxies in which the SFR is regulated by the mass of gas present in a galaxy, and calculated the relation expected when a galaxy evolves in a quasi-equilibrium state.  They argued that this gas regulation would work at least since $z \sim 2$ and possibly also at higher redshifts.  Their derived Z(M,SFR) relation has a particular analytic form and depends on the internal parameters describing the regulator system, specifically the star-formation efficiency $\epsilon = SFR/M_{gas}$ (which is therefore the inverse of the gas depletion timescale $\tau_{gas}$), and the mass-loading of any wind that drives gas out of the system, their $\lambda = outflow / SFR$.  Furthermore, the $Z(M,SFR)$ relation would only evolve to the extent that these two parameters change with epoch (at fixed galactic mass), providing an attractive explanation for the claimed existence of an FMR.   Li13 could represent the local SDSS data from Ma10 with their predicted $Z(M,SFR)$ relation with astrophysically plausible values of $\epsilon$ and $\lambda$ and reasonable mass-dependences of these two parameters.  It should be noted that the analytic approximations employed by Li13 have since been verified by more detailed calculations in \citet{pipino14}.

~~~The form and possible epoch-(in)dependence of the $Z(M,SFR)$ relation thus emerges, along with other measures of the state of high redshift galaxies, as a very powerful diagnostic of the regulation of star-formation in typical MS galaxies. The definition of this relation at high redshifts is therefore a priority.

~~~There are a number of ways to estimate the chemical abundances in galaxies, especially through the analysis of emission line ratios of gas in star-forming HII regions.  Over the years, a number of diagnostics have been developed based on strong lines, including the well-known $\rm R_{23}$ parameter first introduced by  \citet{pag79}.  This is based on the ratio of oxygen lines ${\rm [O\,II]\,}{\lambda\,3727}$  and ${\rm [O\,III]\,}{\lambda\,5007}$ to H$\beta$.  This ratio must be corrected for reddening using the H$\alpha$ to H$\beta$ ratio. A degeneracy between high and low O/H values that exists for most values of $R_{23}$ 
can be broken using, e.g., the [NII]6584/H$\alpha$ ratio, or other means as summarized in, e.g.,  \citet{henry13}. 
Finally, the O/H calibration also varies with ionization parameter, $q$, which may well vary with redshift but which can be estimated from ${\rm[O\,III]}/{\rm [O\,II]\,}$. In our earlier work \citep{maier05}, we extended the basic $R_{23}$ method by simultaneously fitting all of these five lines (\OII, \Hb, \OIIIa, \Ha, and \NII) for metallicity O/H, reddening $A_V$ and ionization parameter $q$.  

~~~Unfortunately, at high redshifts around $z \sim 2$, these five lines  are seen in the atmospheric windows for a limited range of redshifts only, namely at $2.1 < z < 2.5$.   At higher redshifts, the H$\alpha$ and [NII] lines which are critical for the practical implementation of the $R_{23}$ method are shifted redward of the K-band window. Even within this narrow redshift window, the chance of having all five lines clear of strong telluric OH emission is not very high.   In addition, near-infrared (NIR) spectroscopy at these faint levels is difficult, even on 8-m class telescopes.

~~~As a result, none of pioneering studies of metallicities of star-forming galaxies at $z>2$ \citep{petpag04, erb06, maiol08, mannu09} has been based on measurements of all five lines, and only a handful, i.e., four galaxies of Eb06, five of \citet{pettini01}, two galaxies from LSD \citep{mannu09}, and a few AMAZE objects \citep{maiol08} have measurements of even three or four of the lines. The derived O/H metallicities have therefore been based on a series of ad hoc calibrations of more limited line ratios, and often just [NII]/H$\alpha$, as in the large sample of Eb06. These suffer from significant scatter and systematic uncertainties arising from the poor knowledge of  ionization parameter (in case of the N2 calibration)  and/or reddening.  Even now, there are only six  $z>2$ galaxies  with the five emission line fluxes measured in the literature, four of which are gravitationally lensed and all having rather low mass \citep{richard11,belli13,masters14}.

~~~There are therefore several caveats regarding the extensive Eb06  study of metallicities of 87 galaxies at $2.1<z<2.5$.
The ionization parameter could not be measured  for any of these galaxies, and
only coarse oxygen abundances using the [NII]/H$\alpha$ ratio
were determined  by applying the
so-called N2 calibration
\citep{petpag04}, which does not take the ionization parameter  into account.
In Fig.\,\ref{N2_IonPar} we show the [NII]/H$\alpha$ ratio vs. metallicity as a function of the ionization parameter $q$ (dotted lines), based on Fig. 7 in \citet[][KD02 in the following]{kewdop02}.
The reason why the N2 calibration works at low redshift is that ionization is correlated with metallicity. However, if this is also true at higher redshift is a matter of debate, and it is not clear which range of ionization parameters one should adopt at each metallicity at higher redshift.

To derive the MZR at $z>2$, Eb06 used the averaged  [NII]/H$\alpha$ ratios  in six stellar mass bins each containing $\sim 15$ galaxies, and estimated the O/H metallicities of these six averaged spectra using the N2 calibration. The horizontal lines in Fig.\,\ref{N2_IonPar} show the  measurements of [NII]/H$\alpha$ and the inferred range in O/H (the ionization parameter, i.e., the location on a corresponding dotted line, is not known), while the filled blue circles show the O/H measurements derived by Eb06.
The magenta oblique line has been transformed from the N2 calibration to the KD02 calibration (used in this paper) by applying the conversion given in Tab.3 of  \citet{keweli08}. It is obvious from Fig.\,\ref{N2_IonPar} that, without knowledge of the ionization parameter (dotted lines),
the metallicity estimates have very large uncertainties, especially at high metallicities: the six horizontal blue lines of possible O/H at a given [NII]/H$\alpha$ extend over $0.7-1$\,dex.
It should be noted that it was stated by \citet{petpag04} that ``with the N2 calibrator it is possible to estimate the abundance of oxygen to within a factor of 2.5 at the 95 per cent confidence level.'' Factor 2.5 corresponds to about 0.4\,dex, which is comparable to the scatter in O/H at a given log([NII]/H$\alpha$) in  Fig.\,\ref{N2_IonPar} (indicated by the blue lines).

Furthermore, only  4 of the 87 galaxies in the Eb06 sample have [OIII]$\lambda 5007$ and H$\beta$ line fluxes measured in addition to H$\alpha$ and [NII], i.e., no information on type-2 AGN contamination is available for all but 5\% of the galaxies in the Eb06 sample. It is therefore unclear whether the ionization in these objects
is dominated by an AGN, which would lead to highly uncertain metallicity measurements. 
Unfortunately, these uncertainties in both the ionization parameter and AGN contribution are now strong impediments to a clear understanding of these galaxies and of the chemical evolution at $z>2$.

~~~To overcome these uncertainties,  we have carried out NIR follow-up spectroscopy with Subaru-MOIRCS and VLT-SINFONI of 20  star-forming galaxies at $2.1<z<2.5$, selected from the zCOSMOS-deep sample.  As well as providing a large enough parent sample, using zCOSMOS-deep galaxies enables us to also use the wealth of information on these galaxies that is available from the COSMOS survey.

~~~ The aim of this  paper is to  use 
oxygen gas abundances, stellar masses and SFRs for a sample of zCOSMOS galaxies at $2.1<z<2.5$ to study the MZR and FMR at $z>2$. The paper is structured as follows: In Sect. 2 we describe the selection of the targets and the Subaru-MOIRCS and VLT-SINFONI observations. In Sect. 3 we present the measurement of emission line fluxes, and the derivation of
SFRs and metallicities using a representation of the KD02 models. We also discuss the possible contamination by AGNs using the BPT \citep{bpt}  diagram. In Sect. 4 we present the SSFR-mass, MZR and Z(M,SFR) relations at $z\sim 2.3$, and discuss our findings. Finally in Sect. 5 we present our conclusions.

~~~A {\sl concordance}-cosmology with $\rm{H}_{0}=70$ \kmsmpc,
$\Omega_{0}=0.25$, $\Omega_{\Lambda}=0.75$ is used throughout this
paper.  Note that {\sl metallicity} and {\sl abundance} will be taken to denote {\it gas oxygen abundance}
throughout this paper, unless otherwise specified.  In addition, we use dex throughout to denote the anti-logarithm, i.e., 0.3 dex is a factor of 2.


\section{The Data}

\subsection{Parent sample and selection of targets for the NIR spectroscopic follow-up}

~~~ The target galaxies for the NIR spectroscopy with MOIRCS and SINFONI  were selected from the 
zCOSMOS-deep galaxy catalog.  zCOSMOS-deep is measuring several thousand redshifts 
in a broad redshift range $1.4 < z < 3.5$.  
We first selected  a parent sample which we call \emph{ELPAR} of $\sim 800$ zCOSMOS-deep galaxies with redshifts $2.1<z<2.5$ of high confidence classes 2.5, 3 or 4, 
i.e., reliable redshifts \citep[see][]{lilly07}. Class 3 and 4 are very secure redshifts, while class 2.5 are less secure spectroscopic redshifts, but the decimal place .5 indicates that the spectroscopic and photometric redshift of objects with confidence class 2.5 are consistent to within $0.08(1+z)$.
X-ray detected objects were excluded from the sample ELPAR using the XMM \citep{brusa07}, and Chandra COSMOS
observations \citep{elvis09}. 

~~~For the NIR spectroscopic follow-up we only selected those ELPAR galaxies with redshift confidence classes 3 or 4, i.e., the most reliable redshifts based on unambiguous multiple spectral features that leave no room for any doubt about the redshift.
 For each galaxy, we estimated an SFR using the model spectral energy distribution (SED) fits obtained, at the
spectroscopic redshift, by the ZEBRA \citep[Zurich Extragalactic Bayesian Redshift Analyzer,][]{feld06,feld08} code applied to the multi-band COSMOS photometry in the bands u, B, V, g, r, i, z and Ks.
We used this SFR to predict an H$\alpha$ flux by applying the \citet{ken98} conversion of H$\alpha$ luminosity into SFR (cf. Sect.\,\ref{SFRs}).  We then selected only those objects with SED-based SFRs larger than $\sim 30$\msun/yr for the SINFONI observations, corresponding to an expected observed H$\alpha$ flux larger than $\sim 10^{-16}$ergs/s/cm$^{2}$ at these redshifts, but also included targets with lower SED-based SFRs  as fill-in objects during the Subaru MOIRCS mask
preparations.
%


\subsection{MOIRCS Observations}

Multi-object spectroscopic observations were performed with MOIRCS \citep{suzuki08}
in the second half of the nights  of 2010 January 25-28. We used MOIRCS in MOS mode with
the K1300 grism for the K-band, VPH-H grism for the H-band, and VPH-J
grism for the J-band observations.
The spectral resolution with a 0''.8 slit is $R\sim 950 $ in K, and $R\sim 1900 $ in H
and J. Eight main target galaxies at $z>2$ were observed with one slit mask, with a total exposure time of $\sim 6$ hours in each of the bands J, H, and K.   The seeing varied between 0''.4 and 0''.7.

The target acquisition was done using 6 alignment stars, and the telescope was dithered along the slit by 2''.0. The single exposure
times were 1200s. At the end of each observing night, an A0V type star was observed for flux calibration with the same instrument
configuration as the targets, and at different positions on the mask for the VPH grisms, to take into account that the sensitivity curve  shifts for the VPH grisms, when the slits at different x positions (in raw image) are used.
It should be noted that this sensitivity curve dependence of the VPH grisms on the position of the slits was seriously hampering the observations, in terms of restricting a lot the target selection and diminishing the S/N obtained.


\subsection{MOIRCS Data Reduction}
\label{MOIRCSRedu}

The data reduction was done using the MOIRCS MOS Data Pipelines
(MCSMDP) reduction script \citep[for more details see][their
Sect.\,3.1]{yoshi10}. This included flatfielding by a dome flat, rejection of bad pixels and
cosmic-rays, sky-subtraction
between each exposure pair, and distortion correction of the MOS frames.
Each 2-D spectrogram was then cut out based on the mask
design file,  wavelength
calibration was done based on the OH sky emission lines, residuals of
the sky subtraction (due to time variation of OH emission during
dithering observations) were subtracted by fitting with a quadratic
equation along the column, and finally all frames were co-added with
appropriate offsets and weights.

The response spectra, including atmospheric transmission and instrument
efficiency, were derived using
the A0V-type  stars Hip 68713 and Hip 74050.
The data of these standard stars
were reduced to a two-dimensional
spectrum in the same way as for the galaxy targets, and the
continuum was then traced with a polynomial to extract a one-dimensional
spectrum with Iraf:apall. Finally the one-dimensional spectrum of the
standard star was divided by the best-fit model spectrum to obtain the response
spectrum.
The absolute flux calibration was then performed by normalizing to the
J-, H- or K-band total magnitude, respectively.


\subsection{SINFONI Observations}

Observations with the SINFONI integral field spectrograph \citep{eisenh03,bonnet04}
in up to three bands (J, H, and K) were
performed between the beginning of January and the end of April 2010. 
We used ``on-source'' dithering for all targets, which are sufficiently  small to fit in the region of overlap of all science exposures for a target.  Briefly, two successive exposures  are dithered by roughly 1/2 of the field-of-view (by about 4 arcsec), so that the 
frames can be subtracted pairwise for background subtraction. 

Exposure times were 40\,min
for K-band, and 60\,min for J and H-band observations.
The spectral resolution is $R\sim 4000 $ in K, $R\sim 3000 $ in H, and $R\sim 2000 $ in J-band.
The seeing varied between 0''.7 and 1''.2.  The accuracy of the zCOSMOS-deep redshifts  is about $\sigma_z \sim 0.003$ at
$z\sim 2.3$, and they are generally absorption line redshifts from rest-frame ultraviolet (UV) features
(from the VIMOS optical spectroscopy), which 
often show a significant velocity offset with respect to the emission line gas (see columns $z_{UVabs}$ and $z_{EL}$ in Tab.\,\ref{SINF_MOIRCS}).
Since all five emission lines are clear of OH lines for only a limited range of redshifts, our first step was to
observe the H$\alpha$ (and [NII]) lines in K-band for the 12 galaxies given in Tab.\,\ref{SINF_MOIRCS}.  
These measurements allowed us to focus 
the H$\beta$, [OIII] and [OII] observations (in H- and J-band) on those galaxies lying at optimal redshifts 
in terms of the OH emission contamination and emission line flux detections of (most of the) five lines.
Tab.\,\ref{SINF_MOIRCS} indicates the lines that could be observed for each  object.

\subsection{SINFONI Data Reduction}
\label{SINFRedu}

We reduced the SINFONI 
data with the SPREAD software, especially
developed for SPIFFI \citep{schreib04,abut06}.  The
data reduction is analogous to standard procedures applied to
NIR long-slit spectroscopy but with additional processing to
reconstruct the three-dimensional (3D) data cube. The main
reduction steps applied to each science target for a given
instrument band are as described in Sect.\,4.2 of \citet{foeschr09}.  A special step in the reduction is that the
wavelength calibration and sky subtraction steps were repeated
with optimization following the method described by \citet{davies07}.

The data of the standard stars were reduced in a similar way as the science data. Flux
calibration was performed 
using the
broadband magnitudes of the standards. Correction for
atmospheric transmission was done by dividing the science
cube by the integrated spectrum of the standard.

%
\section{Measurements and Derivation of SFRs, Masses and Metallicities}

%
\subsection{Emission Line Fluxes}
\label{emline}

~~~ Emission line fluxes were measured from the calibrated spectra
following the procedure described in \citet{maier05}. The package splot in IRAF was used, and the flux errors were usually dominated by systematic uncertainties in establishing the local continuum.
For each individual galaxy, Tab.\,\ref{SINF_MOIRCS} reports either the emission
line fluxes or their upper limits. All five emission line fluxes could be measured for only a few galaxies because of, e.g., poor seeing during observations in one of the NIR bands, low galaxy surface brightness, the overestimation of some SFRs derived from the SED, and atmospheric OH lines coinciding with the positions of some emission lines, as indicated in the respective columns of Tab.\,\ref{SINF_MOIRCS}.
We note in particular that:

~~~ {\it (i)}  We were able to measure \Ha\, for all 20 galaxies at $2.1<z<2.5$
reported in Tab.\,\ref{SINF_MOIRCS}.

~~~ {\it (ii)} The \NII\, line was measured for three galaxies, affected by an OH line for seven
galaxies, and too faint to be detected for ten
objects. We determined upper limits to the \NII\, flux for the ten $2.1<z<2.5$
galaxies for which  the \NII\, flux was unaffected by a night sky line, and
 Tab.\,\ref{SINF_MOIRCS} lists the
$3\sigma$ upper limits.

~~~ {\it (iii)} \OIIIa\, could be measured for 13 galaxies. For other two galaxies in which the \OIIIa\, line was affected by a strong OH line but \OIIIb\, was not, the \OIIIa\, flux was derived using the theroretical line flux ratio between \OIIIa\, and \OIIIb, by multiplying the \OIIIb\, flux by a factor of three. Three galaxies do not have an H-band spectrum, resulting in no measurement of [OIII] and \Hb, and two galaxies have both \OIIIa\, and \OIIIb\, affected by night sky lines.

~~~ {\it (iv)} We were able to
measure  \Hb\, fluxes for  eight of the $z\sim2.3$ galaxies.
For the other galaxies the \Hb\, line was either too faint to be detected, or coincided with a strong OH
line, or the H-band spectrum was not observed, hence we were unable to  measure the \Hb\, line flux.
In the problematic cases, we assumed that the $\Ha/\Hb$ ratio is in the range 
$2.86-6$, corresponding to typical extinction values of star-forming galaxies of
$0<A_{V}<2.1$. We used these values for \Hb\, 
to get an O/H estimate by
simultaneously fitting all the  lines for 
metallicity O/H, reddening $A_V$ and ionization parameter $q$, as described in \citet{maier05} and in Sect.\,\ref{oxabund}.

~~~ {\it (v)} \OII\, could be measured for ten galaxies, and we obtained $3\sigma$ upper limits for two additional galaxies. Four galaxies do not have an J-band spectrum, implying no measurement of [OII], and for one galaxy the region of the spectrum where [OII] should be detected is affected by a bad pixel area of the MOIRCS instrument. For three galaxies the [OII] line was affected by a strong OH skyline.

To calculate SFRs for the sample we need to know the absolute flux of \Ha.
The SINFONI spectra were integrated in circular apertures with radii typically 1''-1''.5 for these seeing limited data. As mentioned in Sect.\,\ref{SINFRedu}, the SINFONI data were reduced with the same procedure as used by \citet{foeschr09}, so the uncertainties of the absolute flux calibration should be $\sim$ 10\%, as estimated by \citet{foeschr09}. In the good seeing conditions encountered for the MOIRCS observations (seeing between 0''.4 and 0''.7), the resulting emission line fluxes from the MOIRCS spectroscopy should approximate ‘‘total’’ fluxes, and the absolute calibration was also checked versus the total J, H or K magnitude, as described in Sect.\,\ref{MOIRCSRedu}.
Additionally, we find a good agreement between the SFRs derived from \Ha\, from the SINFONI and MOIRCS spectra and the SFRs derived from UV (as described in Sect.\ref{SFRs}), which reinforces the reliability of the absolute calibration of the NIR data.

%
\subsection{Star Formation or AGNs?}

To proceed with the interpretation of the emission line properties, one
has to establish whether the  ionization is of stellar origin, or associated with AGN activity.
To identify galaxies possibly dominated by an AGN, we use the BPT \citep{bpt} 
diagram.
Fig.\,\ref{BPT_z23} shows those zCOSMOS galaxies with NIR spectroscopy for which the [NII]/\Ha\, and [OIII]/\Hb\, ratios  (or upper limits of  [NII] or \Hb) were measured, as given in Tab.\,\ref{SINF_MOIRCS}: 
two galaxies with all four lines measured (filled
magenta circles with error bars), five galaxies with [OIII], \Hb, and \Ha\,
measured and an upper limit of [NII] (filled magenta circles with
arrows indicating the [NII] upper limits), and four galaxies for which \Hb\, could not be measured, but was
estimated using the known \Ha\, flux as described in Sect.\,\ref{emline} (open circles).
Galaxies with 
non-degenerate
oxygen abundances (see Sect.\,\ref{oxabund}) are overplotted as yellow crosses.
\citet{kewley01} combined stellar
population synthesis and photoionization models to build the
first purely theoretical classification scheme for separating pure
AGNs from galaxies containing star formation (dashed black line). The empirical curve of \citet[][solid black line]{kauf03} 
separates star-forming galaxies in the local universe (below/left of the
curve) from AGNs (above/right of the curve).

The [NII]/\Ha\, and [OIII]/\Hb\, ratios have been reported in the literature
for small numbers of galaxies at $z>2$:
only  four galaxies in the Eb06 sample have [OIII]$\lambda 5007$ and H$\beta$ line fluxes measured in addition to H$\alpha$ and [NII], and these are shown as blue squares in Fig.\,\ref{BPT_z23}. They lie between the theoretical curve of \citet{kewley01} and the empirical curve of \citet{kauf03}. The two open triangles are measurements derived from stacked spectra by \citet{yoshi10}, and they also lie in the same region of the BPT diagram as the Eb06 galaxies, and this also applies to the two galaxies from \citet{masters14} shown as blue star symbols.

Most of the 11 zCOSMOS galaxies shown in Fig.\,\ref{BPT_z23} have larger ratios of the collisionally excited to recombination lines than present-day galaxies (black circles).
All six galaxies with large [OIII]/\Hb\, close to the division between SF
and AGNs have high ionization parameters.  As mentioned by \citet{brinchmann08}, unusually high ionization parameters might cause the offset of high redshift galaxies in the BPT diagram compared to local ones.

\citet{brinchmann08} found that local SDSS galaxies with high \Ha\, equivalent widths (a proxy for SSFR) at a given mass, 
are displaced towards larger values of the two emission-line ratios used in the BPT diagram. 
The median SSFR of the MS is higher by a factor of 20 at $z>2$ compared to the local MS  \citep[cf. Eq. 1
in][]{peng10}. This implies that MS galaxies at $z>2$, although  displaced towards larger values of the two emission-line ratios in the BPT diagram, may still be ionized by recent star formation instead of AGN activity. Recent work by \citet{kewley13,kewley13b}  shows that, for the more extreme ISM conditions at higher redshifts (e.g., higher ionization parameters), the star-forming sequence may evolve compared to local galaxies.
In Fig.\,\ref{BPT_z23} the open circles show local galaxies with $z<0.04$ from the Nearby Field Galaxy Sample (NFGS) of \citet{jansen}, and the two dashed blue lines in the left part of Fig.\,\ref{BPT_z23} show the upper and lower boundary of the evolved (compared to local galaxies like the NFGS objects) star-forming sequence at $z \sim 2.3$ using the equations in \citet{kewley13,kewley13b}. Galaxies containing AGNs form a mixing sequence with purely star-forming galaxies, and this mixing sequence is shown by  the two blue dashed lines in the right part of the diagram using the values from Tab.\,1 in \citet{kewley13b} for $z=2.5$.
The position of the  zCOSMOS  $z>2$ galaxies (and of galaxies at $z>2$ from the literature depicted in Fig.\,\ref{BPT_z23}) in the region of the BPT diagram occupied by star-forming galaxies at $z \sim 2.3$ (indicated by the two blue dashed lines on the left) implies that they are unlikely to be dominated by AGNs.

%
\subsection{Star formation rates (SFRs)}
\label{SFRs}

For the 20 zCOSMOS galaxies at $2.1<z<2.5$ with NIR spectroscopy, the \Ha\, luminosities were corrected for extinction using, when available, the $A_{V}$ values computed as described in Sect.\,\ref{oxabund}, or assuming an $A_{V}=0.5$ (roughly the mean value of the galaxies with measured $A_{V}$) when no additional information was available.
These extinction corrected  \Ha\, luminosities were used to calculate the SFRs, by applying  the \citet{ken98} conversion of H$\alpha$ luminosity into SFR: $\rm{SFR} (M_{\odot}\rm{yr}^{-1}) = 7.9 \times 10^{-42}
\rm{L}(\rm{H}\alpha)\rm{ergs/s}$.

The SFRs for $z>2$ galaxies of the ELPAR parent sample were computed from their UV luminosities using the relation based on their 1500 \AA\, luminosity, following Eq.\,5 of \citet{daddi04}, namley: $logSFR_{UV}=logL_{1500corr}-27.95$, for a  \citet{salp55} IMF.
The extinction correction was done using Eq.\,4 of \citet{daddi04}, i.e., $E_{(B-V)_{UV}}=0.25(B-z_{mag}+0.1)$.
The procedure is described in detail in Sect.\,5.1.3 of \citet{mancini11}.  
Using a sample of zCOSMOS galaxies at $z>2$ with SINFONI NIR spectroscopy of H$\alpha$ and with a comparable range of stellar masses and SFRs as the galaxies studied here,
\citet{mancini11} compared the SFRs inferred from UV and from  H$\alpha$, and showed this UV calibration of the SFR to be reliable at $z>2$ (see, e.g., their Fig. 14).

Eb06 used the same value of E(B-V) for the stellar UV continuum and for the nebular emission
lines when calculating their published SFRs. This is in agreement with several studies who found similar extinction for the stellar continuum and nebular emission \citep[e.g.,][]{cowbar08,hainl09,woffo13}. However, the relation between stellar and nebular exinction is uncertain, and other studies found that nebular regions are more extincted than the stellar continuum.
\citet{mancini11} found that, when the additional dust extinction caused by nebular regions is calculated using \citet{calz94,calz00}, there is a closer agreement between the \Ha-derived SFRs and other SFR indicators for $z>2$ galaxies.
Therefore, in this paper we re-calculate the (S)SFRs of the $z>2$ Eb06 sample from the published \Ha\, fluxes of \citet{erb06b}, assuming $A_{V,HII} = A_{V,SED}/0.44$, following \citet{calz94,calz00} who found that the Balmer emission lines of HII regions are more extincted than the UV stellar emission 
by this amount.
In addition, in agreement with this prescription, \citet{foeschr09} also found that nebular
regions are about two times more extincted than the stellar
continuum in their SINS sample of star-forming galaxies at
$z>2$.


\subsection{Stellar Masses}

The optical to IRAC photometric data from COSMOS were used to perform an SED fitting analysis 
to estimate the galaxy stellar mass. The \emph{HyperzMass} software was used, which is a modified version of the public photometric redshift code \emph{Hyperz} \citep{bolzo00}, and details of the method are described in Sect.\,2.3 of \citet{bolzo10}.
Briefly, \emph{HyperzMass} fits photometric data points with the \citet{bruzcharl03} synthetic stellar population models, and picks the best-fit parameters by minimizing the $\chi^{2}$ between observed and model fluxes. The best-fit results depend strongly on the assumptions regarding initial mass function (IMF), metallicity, stellar population models, extinction law, and star formation history (SFH). 
The stellar mass is obtained by integrating the SFR over the galaxy age, and correcting for the mass loss in the course of stellar evolution.

The choice of the different parameters was discussed and compared by \citet{bolzo00} for the $z<1.5$ redshift regime,  and by \citet{mancini11} for the $z>1.5$ redshift range, so we refer the reader to those papers for details.
Specifically, 
\citet{mancini11} showed (see their Sect. 3) that COSMOS galaxy stellar masses at $z>1.5$ derived from SED fitting can be considered reliable within  $\sim 0.2-0.3$\,dex, i.e., within a factor $1.5-2$.

The stellar masses of the Eb06 sample are the integral of the SFR over the lifetime of the galaxy (without correction for the mass loss) and so represent the total mass of stars formed rather than the mass in living stars at the time of observations. They have been converted into the living stellar mass by assuming the usual instantaneous return to the ISM of a fraction of the mass converted into stars. We chose this fraction to be $R=0.4$ like in Li13, based on \citet{bruzcharl03} stellar population models.
We note that the stellar mass calculated in this paper is always  the actual mass of long-lived stars, and when  literature data is used we always correct to this definition if necessary.

%
\subsection{Oxygen abundances}
\label{oxabund}
%

For the zCOSMOS $2.1<z<2.5$ galaxies we derived values of O/H and  dust extinction  $A_{V}$ using the
approach described in \citet{maier05}.
This adopts the models of KD02, who developed a set
of ionization parameter and oxygen abundance diagnostics based on the use of
strong rest-frame optical emission lines.  The method of \citet{maier05} performs a
simultaneous fit to all available emission line fluxes  in terms of extinction
parameter $A_{V}$, ionization parameter $q$, and O/H, and is 
described in detail in Sect.\,3 of \citet{maier05}.
An upper limit to [NII]$\lambda 6584$ often suffices to break the $R_{23}$ degeneracy and allows O/H determination based on this
$\chi^{2}$ analysis, as shown in \citet{maier05, maier06}.
It should be noted that the point of the line-fitting approach developed in \citet{maier05} is to use all of the data and the best available emission line models (in this case KD02) without applying other priors that may or may not be valid.

For the reasons discussed above in Sect.\,\ref{emline}, 
the fluxes of all five emission line fluxes could  be reliably measured for only a few of the 20 observed 
$z>2$ galaxies. 
When \Hb\, was undetected, we estimated the  \Hb\, flux as described in Sect. \ref{emline} 
(see also footnotes in Tab.\,\ref{SINF_MOIRCS}).
We do not correct for the possible effect of Balmer stellar absorption lines on the measured line
fluxes, because of the high equivalent widths (EWs) of the emission lines of the galaxies studied. For a wide range of star formation histories and for plausible IMFs, the stellar absorption at \Ha\, is always $<$5\AA\,
\citep[see, e.g.][]{brinchmann04}, small compared to the EWs of the emission lines in our studied galaxies.
The stellar absorption lines corrections are typically larger for the \Hb\, fluxes, 
but the \Hb\, line fluxes of the galaxies in our sample have quite large error bars, so this correction should be small compared to these uncertainties.

The error bars of O/H and $A_{V}$ that we report in  Tab.\,\ref{SINF_MOIRCS}
are the formal 1$\sigma$ confidence intervals for the projected best-fitting values.
As described in Sect.\,3.3 of \citet{maier05}, the error bars of the oxygen abundance are
given by  the range of oxygen abundance for those 
models with $ \chi^{2}$ in the range  $\chi_{min}^{2} \leq \chi^{2} \leq
\chi_{min}^{2} +1 $ (where  $\chi_{min}^{2}$  is the minimum $\chi^{2}$
of all allowed models for a given galaxy), corresponding to  a confidence level of 68.3\%
for one single parameter. 
We consider a
  possible second solution corresponding to a
  second (higher)  peak,  only if the minimum $\chi^{2}$ of the higher peak
  lies in the range  $\chi_{min}^{2} \leq \chi^{2} \leq \chi_{min}^{2}
  +1 $ \citep[cf. Fig.\,4 in][]{maier05}.
Our $\chi^{2}$ analysis takes into account all emission line fluxes measured and their error bars, and 
it turns out that only seven out
of the 20 $z>2$ galaxies studied (35\% of our sample) get a non-degenerate O/H solution with the achieved quality of the data.

%
\section{Results and Discussion}


\subsection{The SSFR$-$mass relation  at $2.1<z<2.5$}
\label{SSFRMass}

Before discussing the MZR relation at $z \approx 2.3$, it is useful to consider the location of the $z>2$ galaxies in the SSFR$-$mass diagram. 
\citet{peng10} linked the SSFR measurements at $z\sim 2$ and $z\sim 1$ of \citet{daddi07} and \citet{elbaz07} with the zCOSMOS data and the SDSS relation to derive an equation of the evolution of the SSFR as a function of mass and time (their Eq.1). The dashed black (almost horizontal) lines in Fig.\,\ref{SSFRMass_zgt2} show the SSFRs calculated using their Eq.1
for $z=2.1$ 
and $z=2.5$, assuming a weak dependence of SSFR on mass (as observed for local SDSS galaxies), SSFR$\propto$M$^{\beta}$ with $\beta=-0.1$. \citet{peng10} and others showed that the local MS relation has a dispersion of $\sim 0.3$\,dex about the mean relation, which is indicated by the dotted (almost horizontal) lines in Fig.\,\ref{SSFRMass_zgt2}.

The SFRs for the individual zCOSMOS galaxies with NIR follow-up (larger magenta symbols in Fig.\,\ref{SSFRMass_zgt2}) were derived from \Ha, while for the parent ELPAR sample (cyan circles) we used SFRs derived from UV data as described in Sect.\,\ref{SFRs}.
We used the \Ha\, values of the individual galaxies from Tab.\,1 in \citet{erb06b} to derive SSFRs as described in Sect.\,\ref{SFRs} for the Eb06 galaxies. 
The oblique dashed lines show SFRs of 10\,\msun/yr and 30\,\msun/yr, respectively, which are roughly the selection thresholds of the zCOSMOS ELPAR (cyan filled circles) and Eb06 sample (blue filled circles).

Because of these SFR selection thresholds, when calculating the median (S)SFRs at a given mass, 
the incompleteness at lower masses \citep[c.f.][]{stringer11} produces  higher median SSFR values at lower masses than the relation from \citet{peng10}, because low mass lower SFR galaxies are missed from the sample (Fig.\,\ref{SSFRMass_zgt2}). This is shown in Fig.\,\ref{SSFRMass_zgt2} by the filled blue squares which indicate the mean SSFRs values in the  mass bins used by Eb06. 
We therefore defined  MS objects as galaxies in the region delineated by Eq.1 in \citet{peng10},
i.e., the area between the dotted lines in Fig.\,\ref{SSFRMass_zgt2}.
We also note that massive galaxies which are highly obscured by dust are probably missed from both the Eb06 and the zCOSMOS samples, as indicated by the lower SSFRs at the high mass end in Fig.\,\ref{SSFRMass_zgt2}. This was also noted and discussed in Sect.3.3 of \citet{mancini11}. Since it affects both the Eb06 and the zCOSMOS samples, this effect should not affect our findings about the difference in metallicities at higher masses between the two samples.
%


\subsection{The mass-metallicity relation (MZR) at $z>2$}
\label{MZR}

In Fig.\,\ref{MassOH} we compare the MZR of  galaxies at $2.1<z<2.5$ with the relation of SDSS. 
For all galaxies studied, the metallicity O/H was derived from NIR spectroscopy of at least three emission lines, apart from the Eb06 sample, whose metallicities were based on only \Ha\, and [NII].
For the local MZR, we used the median and $1 \sigma $ values of SDSS galaxies from Tab.\,3 of \citet{trem04}, and converted their stellar masses to a \citet{salp55} IMF and their oxygen abundances to the KD02 calibration  by applying the conversion given in Tab.\,3 of  \citet{keweli08}.
The 16th and 84th percentiles, and the medians (50th percentiles) of the distribution of O/H in the respective mass bin are shown as black thick lines for SDSS.
The three thinner black lines show the median SDSS MZR, 
shifted downward by 0.3, 0.5 and 0.7 dex, respectively. Our SINFONI and MOIRCS based O/H measurements for the seven galaxies with non-degenerate
 metallicity measurements are shown as  magenta filled circles. 
Literature data of four gravitationally lensed galaxies  at $2.1<z<2.5$ \citep{richard11,belli13}, with O/Hs converted to the KD02 calibration by applying the conversion given in Tab.\,3 of  \citet{keweli08}, are depicted as black filled circles. 
Analysing their 3D-HST NIR spectroscopy data, \citet{cullen13} used [OII], \Hb\, and [OIII] emission lines in stacked galaxy spectra to measure O/H metallicities at $z>2$; these data are shown in Fig.\,\ref{MassOH} as black open circles.
The MZR of Eb06, with O/Hs from the N2-method \citep{petpag04} converted to the KD02 calibration, is shown in blue.

%
The O/H metallicities of zCOSMOS galaxies and most  literature values obtained using at least three emission lines are systematically  lower by a factor of three to five, at a given mass, than the SDSS relation. On the other hand,  the [NII]/\Ha-based metallicities from stacked spectra  of the work by Eb06 are lower by a factor of two compared to SDSS.
This discrepancy can be due to: (i) the different selections of the samples; (ii) type-2 AGN contamination in the Eb06 sample;  and/or (iii) different metallicity calibrations used.


The first point (i) can be addressed using Fig.\,\ref{SSFRMass_zgt2}:  both the ELPAR (and the zCOSMOS galaxies with NIR follow-up) and the Eb06 samples are quite representative of MS galaxies of relatively high mass. At lower masses $log(M/M_{\odot})<10$, ELPAR galaxies are more representative of MS galaxies, while Eb06 galaxies lie predominantly above the MS, which could bias them towards {\it lower} metallicities. However, the Eb06 metallicity measurements are too high compared to zCOSMOS and the \citet{cullen13} sample, and correcting a bias towards lower metallicity, by making the Eb06 metallicities higher in their lowest two mass bins, would even increase the discrepancy at lower masses.


Regarding the second point (ii), we note that only  four out of 87 galaxies in the Eb06 sample have [OIII]$\lambda 5007$ and H$\beta$ line fluxes measured in addition to those of H$\alpha$ and [NII]. This means that  no information on type-2 AGN contamination is available for 83 galaxies. 
Recent work on the AGN fraction at high redshift \citep[e.g.,][]{trump13} claims the existence of a non-negligible fraction of nuclear activity in higher mass galaxies at $z>1$.
If any type-2 AGNs were among the spectra stacked by Eb06 to measure metallicity, this would have systematically increased their measured [NII]/\Ha\, ratios and hence their derived O/H metallicities, when using the N2 calibration of \citet{petpag04}.


Related to the third point (iii), as mentioned above, the oxygen abundances of Eb06 have been converted from the \citet{petpag04} to the KD02 calibration  by applying the conversion given in Tab.\,3 of  \citet{keweli08}. 
To verify if the \citet{keweli08} conversion, which was calibrated for local galaxies, works at $z>2$, a larger sample of galaxies with five emission lines measured is needed at $z>2$.
The ionization parameter dependence of the N2-method is addressed in both Sect.\,\ref{intro} and Fig.\,\ref{N2_IonPar}, and may be an important drawback of the analysis of Eb06 based on the N2-method.
\citet{newman14} indeed found that the MZR at $z\sim 2$ determined using  the N2-method might be  $2-3$ times too high in terms of metallicity, when the effects of both photoionization and shocks are not taken into consideration.


Assuming that the \citet{keweli08} calibration really holds at $z>2$,
and that the Eb06 metallicities are not strongly affected by either type-2 AGN contamination or the ionization parameter, this would imply that the gas-phase metallicities of galaxies at $z>2$ are between one half and one fifth of those of nearby galaxies of similar mass.
However, we think that the N2-method issues (including the ionization parameter dependence)  and the type-2 AGN contamination are serious issues influencing the Eb06 metallicity measurements.

\subsection{The Z(M,SFR) relation at $z>2$}
\label{ZMSFR}
~~~ Ma10 proposed the existence of an FMR based on comparing SDSS measurements with a limited range in SFR and stellar mass, and higher redshift data which have higher values of SFR that are not observed in SDSS ($SFRs>10\,M_{\odot}/yr$).  The claim that the Eb06 $z>2$ galaxies follow the same FMR as SDSS galaxies 
is therefore based on an extrapolation because of the much higher median SSFR of the MS at $z>2$ compared to the local MS  \citep[cf. 
][]{peng10}.
Eqs.\,2, 4 and 5 of Ma10 are used to predict the O/Hs of high redshift galaxies following a redshift independent Z(M,SFR) given their SFR and masses, and Ma10 claimed that these metallicities are observed up to $z \sim 2.5$ (but not at $z>3$), i.e., that a redshift independent Z(M,SFR) holds.


For the convenience of the reader, we reproduce here Eq.\,2 of Ma10:
\begin{equation}
\label{eq:fit}
\begin{array}{rl}
12+log(O/H)=&8.90+0.37m-0.14s-0.19m^2\\
            &+0.12ms-0.054s^2\\
\end{array}
\end{equation}
where $m$=log(M)--10 and $s$=log(SFR) in solar units.
Because Eq.\,5 of Ma10 is quite similar to their Eq.\,4, we will not show their Eq.\,4 for the comparison in the following and focus only on Eq.\,5, which was reproduced as Eq.1 in Sect.\,\ref{intro}.
We also reproduce Eq.\,40 of Li13:
\begin{equation}
Z_{eq}=Z_{0}\\
+y/(1+\lambda(1-R)^{-1}+\epsilon^{-1}((1+\beta-b)SFR/M-0.15))
\end{equation}
where $Z_{eq}$ is the equilibrium value of the metallicity, $Z_{0}$ is the metallicity of the infalling gas, $y$ the yield, $R$ the returned fraction, $\lambda$ the mass-loading factor, $\epsilon$ the star formation efficiency.  The values of the different parameters $\epsilon$, $\lambda$, $y$, $b$ are given in Tab.\,1 of Li13.


~~~~ It is clear from the above  that whether or not the high redshift galaxies fall on a non-evolving FMR depends on how the local Z(M,SFR) relation is extrapolated. We therefore first explore how the non-evolving prediction up to $z\sim 2.5$ 
 differs when using the original formulations of the local Z(M,SFR) of Ma10 compared to the more physically-based formulations in Li13.  By construction, these should agree in the part of the (M,SFR) plane that is well-sampled by SDSS local galaxies.    We created a grid of oxygen abundances as a function of SFR and stellar mass, with log(SFR/M$_{\odot}$/yr) ranging from -2.0 to 3.0 in 51 steps of 0.1 dex, and log(M/M$_{\odot}$) ranging from  8.0 to 12.0 in 41 steps of 0.1 dex.
At every (SFR, mass) grid point we then calculate the expected oxygen abundance using Eqs.\,2 and 5 from Ma10 and using Eq.\,40 of Li13 with a redshift-independent $\epsilon(m)$, i.e., using in each case a non-evolving Z(M,SFR) but differing in the form of the relation.
Because Ma10 uses a \citet{chabrier03} IMF, 
we converted the masses and SFRs to a \citet{salp55} IMF (as used in this paper) after computation of the expected O/Hs, by adding 0.23\,dex in mass and SFR, respectively. 
The results are shown in Fig.\,\ref{FMRMa10Li13},  the SSFR - mass relation, 
color-coded according to the expected metallicities from the different Z(M,SFR) formulations of Ma10 and Li13.

As discussed in Li13, the form of the high redshift $Z(M,SFR)$ relation will depend on whether the parameters of the regulator system change with time. In particular, there are theoretical reasons to suppose that the star-formation efficiency (their $\epsilon$ parameter) might scale roughly as $(1+z)$ if the gas depletion timescale scales as the dynamical time of the system.  Observationally, this is an open question.   In the spirit of investigating an epoch independent $Z(M,SFR)$, we here consider henceforth the case that $\epsilon$ is epoch-independent, i.e., the case represented by the dashed lines in Fig.\,7 of Li13.

The black open circles show the parameter space in SFR and mass for which the predicted O/Hs from Li13 and Ma10 differ by less than 0.05\,dex.
Not surprisingly, this roughly corresponds to the range in observed SFR ($SFR<10\,M_{\odot}/yr$) and mass values of SDSS galaxies, from Tab.\,1 in Ma10, which were both used to fit the parameters in Eq. 40 of Li13 and to derive Eqs. 2 and 5 of Ma10.
It is obvious from Fig.\,\ref{FMRMa10Li13} that \emph{the extrapolation of  Eq. 40 of Li13 and Eqs. 2 and 5 of Ma10 outside the  SFR-mass parameter space area that was covered by SDSS is rather different.}
As an example, this can be seen in the \emph{grid} color (expected O/H) for a typical star-forming massive galaxy at $z>2$ with log(M/M$_{\odot})\sim 10$ and $log(SSFR/Gyr^{-1})=0.5$, as indicated by the crossing point of the magenta lines in the four panels: the expected O/H decreases from panel d) to b) to a) to c), as indicated by the green, blue and cyan colors of the grid points at the respective crossing point of the magenta lines.


Fig.\,\ref{SFRMass_Obs} shows the SSFR-mass diagrams with the grid color-coded in metallicities like in Fig.\,\ref{FMRMa10Li13}, but for a smaller range in mass and SSFRs corresponding to the range typically covered by galaxies at $z>2$.
The galaxies with NIR spectroscopy at $z>2$ are overplotted: the seven zCOSMOS galaxies with 
non-degenerate
metallicities (filled circles), the Eb06 sample (filled triangles),  and the \citet{cullen13} sample (filled squares).
All data points are color-coded according to their metallicities (e.g., green color for $8.6<O/H<8.9$, c.f. Fig.\,\ref{FMRMa10Li13}), to explore if the measured metallicities (color of the \emph{data} points) and the expected metallicity assuming  Z(M,SFR) is redshift independent (color of the \emph{grid} points) are in agreement.

At a given mass, e.g., at log(M/M$_{\odot})\sim 10.5$, lower metallicities (blue data points in Fig.\,\ref{SFRMass_Obs}) are mainly observed for higher (S)SFRs, in contrast to higher metallicities (green data points) for lower (S)SFR. This explicitly confirms, albeit with only a few objects, that the \emph{SFR is still a second parameter in the mass-metallicity relation at $z>2$}. 

Whether the Z(M,SFR) relation evolves or not compared to $z=0$, i.e., whether it is ``fundamental'', depends on the extrapolation used.  As in Fig.\,\ref{SSFRMass_zgt2}, MS galaxies are assumed to lie in the region between the black dashed/dotted (almost horizontal) lines in the panels of Fig.\,\ref{SFRMass_Obs}. 
The black circles in all panels roughly indicate, as explained above, the highest (S)SFRs observed, at a given mass, for SDSS galaxies.
 It is obvious that all $z>2$ galaxies have higher SFRs ($SFRs>10\,M_{\odot}/yr$) than the SDSS galaxies. This means that \emph{trying to address the question if the local $Z(M,SFR)$ relation holds at $z>2$, i.e., if it is truly an FMR, depends on the extrapolation used.} 

The zCOSMOS $2.1<z<2.5$ galaxies with metallicities measured in this study have log(M/M$_{\odot})>9.8$, i.e., they lie to the right of the magenta line in each panel of Fig.\,\ref{SFRMass_Obs}. 
Using the Ma10 Z(M,SFR) extrapolation 
shown in panel d, a typical MS star-forming massive galaxy at $z\sim 2.3$ with log(M/M$_{\odot})>9.8$ cannot have O/H$<8.6$ if it follows a non-evolving FMR: for such MS galaxies, in the region between the dashed/dotted lines,  the grid points of the predicted FMR in the panel d are green and red, i.e., only  O/H$>8.6$ are expected if the FMR holds at $z>2$ following Eq.5 of Ma10.
Lower metallicities (blue/cyan grid points) are however predicted for MS $z\sim 2.3$ galaxies up to log(M/M$_{\odot})\sim 11$ 
by Li13  (panels a and c).  O/H$<8.6$ values (blue grid points) are also predicted by Eq.\,2 in Ma10 for MS galaxies with lower masses, log(M/M$_{\odot})<10.1$ (see panel b).
Therefore, the $z\sim 2.3$ zCOSMOS galaxies and the \citet{cullen13} data points, which almost all have $O/H<8.6$, 
are actually quite in agreement with the predictions for a non-evolving FMR based on the Li13 formulation.
The metalllicities of some of the zCOSMOS galaxies are in agreement with the extrapolation from Eq. 2 of Ma10, especially at lower masses, 
 but they are not generally in agreement with the FMR predicted by the  extrapolation from Eq. 5 of Ma10. This is shown in panel d by the disagreement between the blue/green color of most zCOSMOS data points and the green/red grid points.

For the  Eb06 sample, we do not show the lowest mass bin, which only has an upper limit for the metallicity.
The other five mass bins all have measured $O/H>8.6$ (green and red filled triangles).
The measured metallicities in these five mass bins  agree with the non-evolving predictions of Eq.\,5  of Ma10, as shown in panel d by the agreement between the color of the grid points and the color of the data points (triangle symbols). However, the Eb06 data agree with the (non-evolving) predictions
of Li13 or Eq. 2 of Ma10 (panel b) only for the highest mass bins.
This means that an FMR based on SDSS and  Eb06 data, as claimed by Ma10, only exists when considering the particular form for the extrapolation represented by Eq. 5 of Ma10.

In general, it should be noted that any $O/H<8.6$ metallicity measurement for a massive (log(M/M$_{\odot})\sim 10$) $z>2$ galaxy will be considered as an ``outlier'' from a non-evolving FMR that is based on Eq. 5 (or the similar Eq. 4) of Ma10.
This  also applies to the metallicities of the \citet{maiol08} and \citet{mannu09} $z>3$ samples, which were therefore claimed by Ma10 to need an evolving Z(M,SFR) relation. On the other hand, using the Li13 extrapolation, such $z>3$ galaxies with low metallicities can be consistent with a non-evolving FMR.

The basic point is that the metallicity data of future larger samples of galaxies at $z>3$ that will soon become available with NIR spectroscopy should be compared to both the Ma10 and Li13 predictions in order to establish whether the FMR evolves or not at $z>3$.  The implied evolution, or lack of it, in the Z(M,SFR) relation will usually depend on the extrapolation of the local relation to higher SFRs.


\section{Conclusions}

\begin{figure}[h]
\includegraphics[width=12cm,angle=270,clip=true]{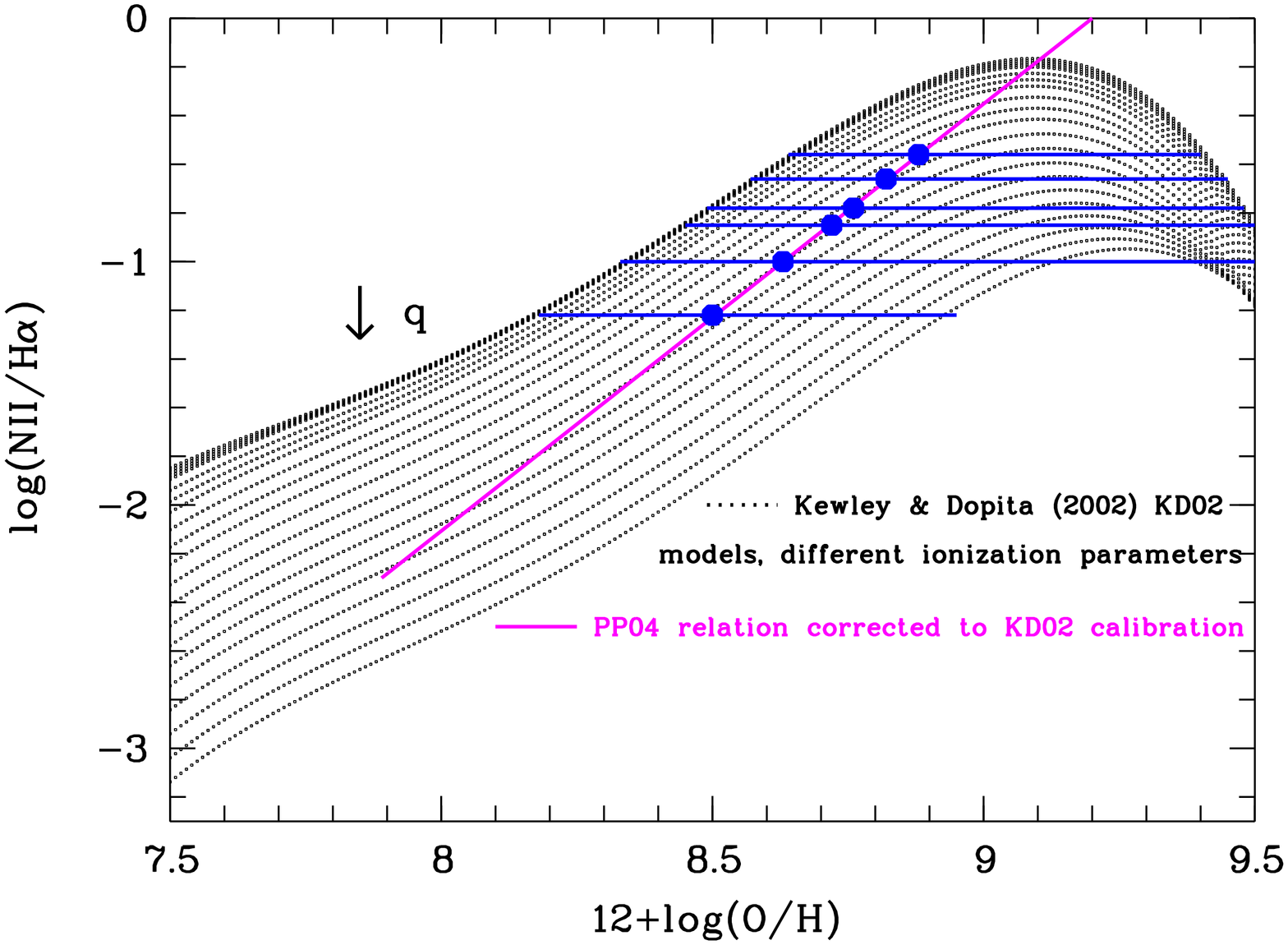}
\caption
{\label{N2_IonPar}
\footnotesize 
The [NII]/H$\alpha$ ratio vs. metallicity as a function of the ionization parameter $q$ (different dotted lines indicate different ionization parameters) based on Fig.\,7 in KD02.
To derive the MZR at $z>2$, Eb06 used the averaged  [NII]/H$\alpha$ ratios  in six stellar mass bins
and estimated the O/H metallicities of these six averaged spectra using the N2 calibration \citep{petpag04}. The horizontal lines indicate the  measurements of [NII]/H$\alpha$ and the inferred range in O/Hs,  
while the filled blue circles show the O/H measurements derived by Eb06, transformed to the KD02 calibration framework.
 It is obvious  that, without knowledge of the ionization parameter (dotted lines), 
the metallicity estimates get very large error bars, especially at high metallicities: the six horizontal blue lines of possible O/Hs at a given [NII]/H$\alpha$ extend over $0.7-1$\,dex.
}
\end{figure}

\begin{figure}[h]
\includegraphics[width=14cm,angle=270,clip=true]{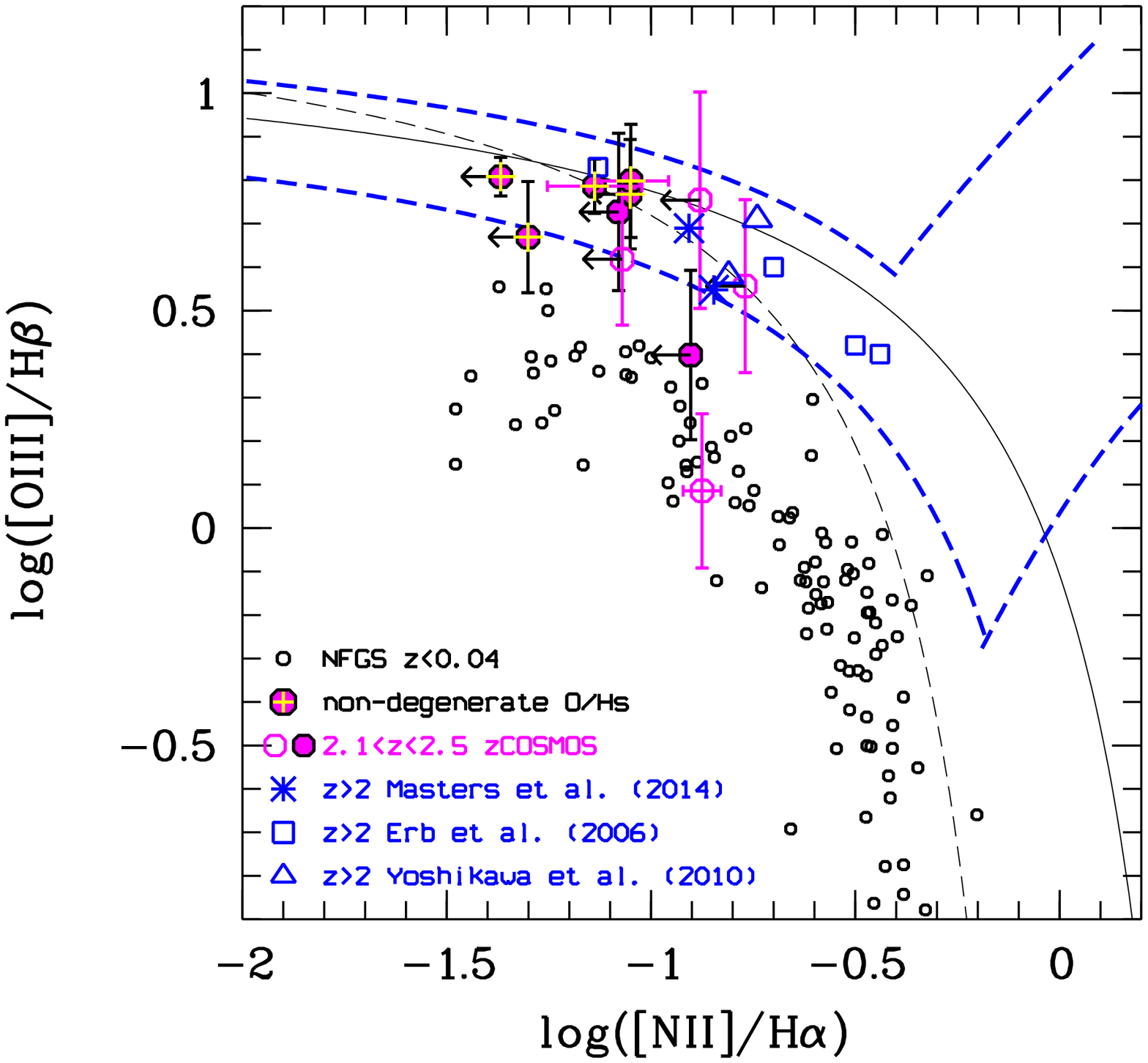}
\caption
{\label{BPT_z23}
\footnotesize 
The BPT \citep{bpt} diagram.
zCOSMOS galaxies with SINFONI and MOIRCS NIR spectroscopy are shown, for which the [NII]/\Ha\, and [OIII]/\Hb\, ratios  or upper limits of  [NII] were measured (arrows) or \Hb\, was estimated from \Ha\, as described in Sect.\,\ref{emline} (open circles).
Galaxies with non-degenerate  oxygen abundances (see Sect.\,\ref{oxabund}) are additionally overplotted with yellow crosses.
The theoretical curve of \citet[][dashed black line]{kewley01} and the empirical curve of \citet[][solid black line]{kauf03}, which
separate star-forming galaxies in the local universe (below/left of the
curves) from AGNs (above/right of the curves) are shown.
Both the [NII]/\Ha\, and [OIII]/\Hb\, ratios have been reported in the literature
for small numbers of galaxies at $z>2$ (blue open squares, triangles and stars).
Recent work by \citet{kewley13,kewley13b}  shows that, for more extreme ISM conditions at higher redshifts (e.g., higher ionization parameter), the star-forming sequence may evolve compared to local galaxies like the NFGS sample \citep{jansen}, as shown by the dashed blue lines in the left part of the diagram (normal star-forming galaxies at $z \sim 2.3$ should lie between the two blue lines).  This implies that all  zCOSMOS $z>2$ galaxies 
depicted in the figure are unlikely to be dominated by AGNs.
}
\end{figure}

\begin{figure}[h]
\includegraphics[width=12cm,angle=270,clip=true]{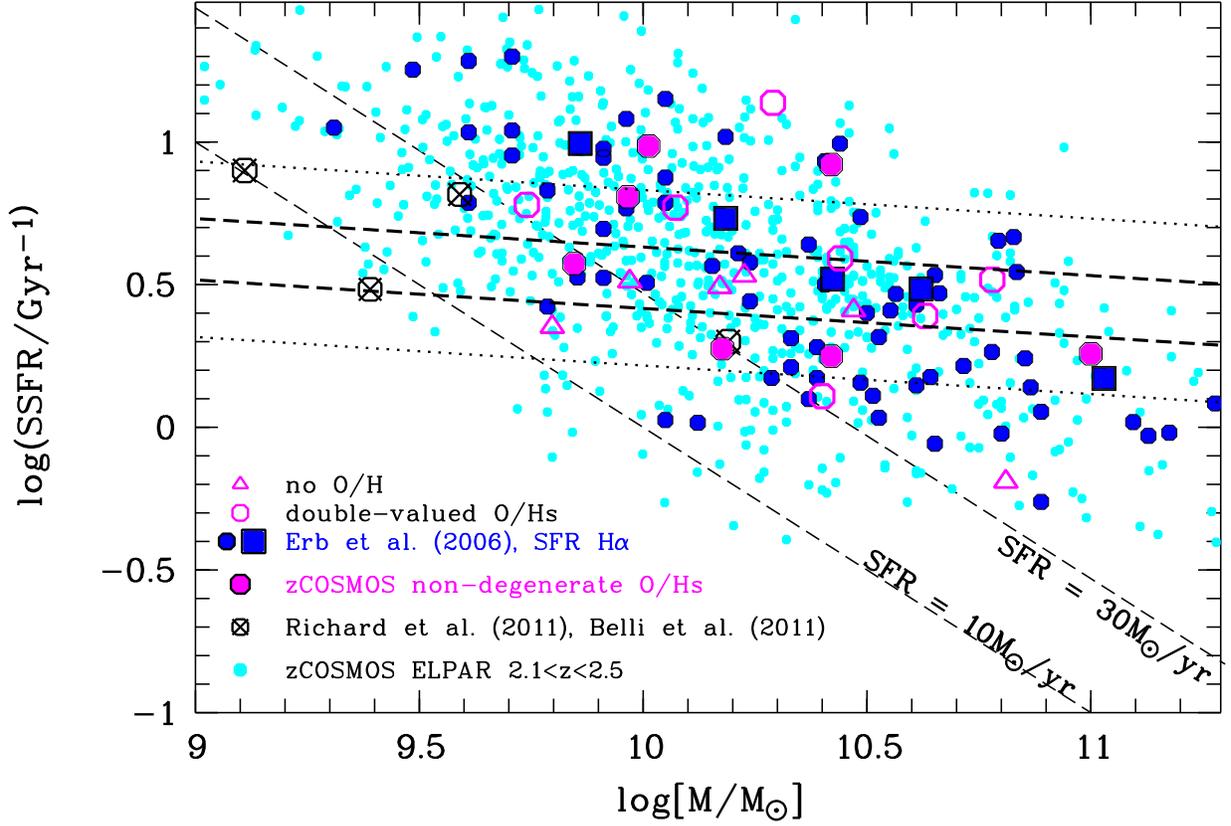}
\caption
{\label{SSFRMass_zgt2}
\footnotesize 
The observed SSFR$-$mass relation  for galaxies at $z>2$.
SFRs for the individual zCOSMOS galaxies with NIR follow-up (magenta symbols) were derived from \Ha, while for the parent zCOSMOS sample ELPAR (cyan circles) we used SFRs derived from UV as described in Sect.\,\ref{SFRs}.
The larger filled magenta circles denote the 7 galaxies with non-degenerate  metallicities, the 7 big open circles are for galaxies with double-valued metallicities, and the 6 open triangles for the zCOSMOS galaxies without an O/H estimate.
The parent sample ELPAR, the zCOSMOS galaxies with NIR folllow-up, and the Eb06 galaxies (blue circles for individual measurements, blue squares for the mean values) are quite representative for MS galaxies (between dashed/dotted, almost horizontal lines), especially at the higher masses.
}
\end{figure}

This study of the MZR and FMR at $z>2$ differs from other studies  by carefully  addressing following issues:  (i) a 
metallicity calibration using  \R23\, based on up to five lines rather than the [NII]/\Ha\, ratio which has a strong ionization parameter dependence (see Fig.\,\ref{N2_IonPar}); (ii)  the location of the $z>2$ galaxies in the BPT \citep{bpt} diagram, which contains information about the conditions of the ISM at these redshifts (see Fig.\,\ref{BPT_z23}); (iii) the discussion of the selection of the samples using the (S)SFR - mass diagram (see Fig.\,\ref{SSFRMass_zgt2}); and (iv) noticing the different formulations of the FMR based on local galaxies data which give different extrapolations at $z>2$ (see Fig.\,\ref{FMRMa10Li13}).

The main results can be summarized as follows.

1. The zCOSMOS galaxies at $z>2$ studied here are not dominated by AGNs.

\begin{figure}[h]
\includegraphics[width=12cm,angle=270,clip=true]{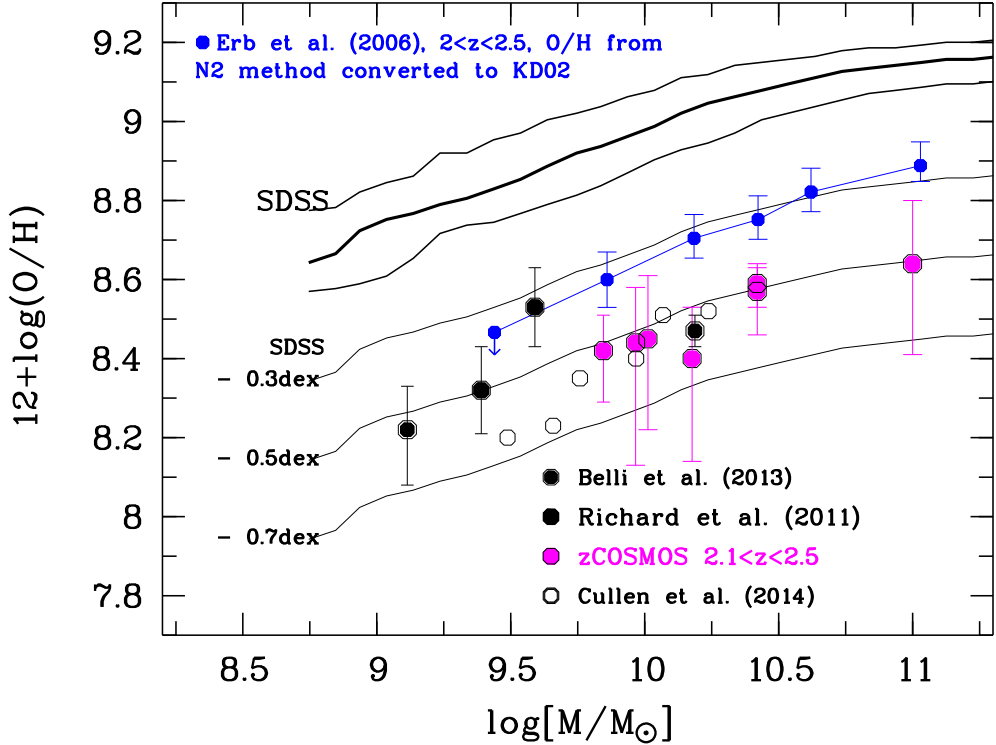}
\caption
{\label{MassOH}
\footnotesize
The MZR of  $2.1<z<2.5$ galaxies, compared to SDSS. For all galaxies O/Hs were derived based on NIR spectroscopy of at least three emission lines, except for the Eb06 data (filled blue circles).
Our SINFONI and MOIRCS O/H measurements are shown as  magenta circles. 
Literature data of four gravitationally lensed galaxies at $2.1<z<2.5$ \citep{richard11,belli13}, 
are depicted as black  filled circles. 
Metallicities from stacked spectra of galaxies with [OII], \Hb\, and [OIII] lines measured, from a recent work by \citet{cullen13} at $z \sim 2.3$, are shown as black open circles.
The three thinner black lines show the median local MZR 
of \citet{trem04}, shifted downward by 0.3, 0.5 and 0.7 dex, respectively. The metallicities of zCOSMOS objects and most galaxies from literature with at least three lines measured are shifted to lower metalllicities, at a given mass, by a factor of three to five compared to SDSS, while  the [NII]/\Ha-based metallicities from stacked spectra  of the work by Eb06 are lower by a factor of two compared to SDSS.
This can be due to the type-2 AGN contamination in the Eb06 sample and/or problems with the N2-metallicity calibration method used by Eb06
(see discussion in Sect.\,\ref{MZR}).
}
\end{figure}

\begin{figure}[h]
\includegraphics[width=14cm,angle=0,clip=true]{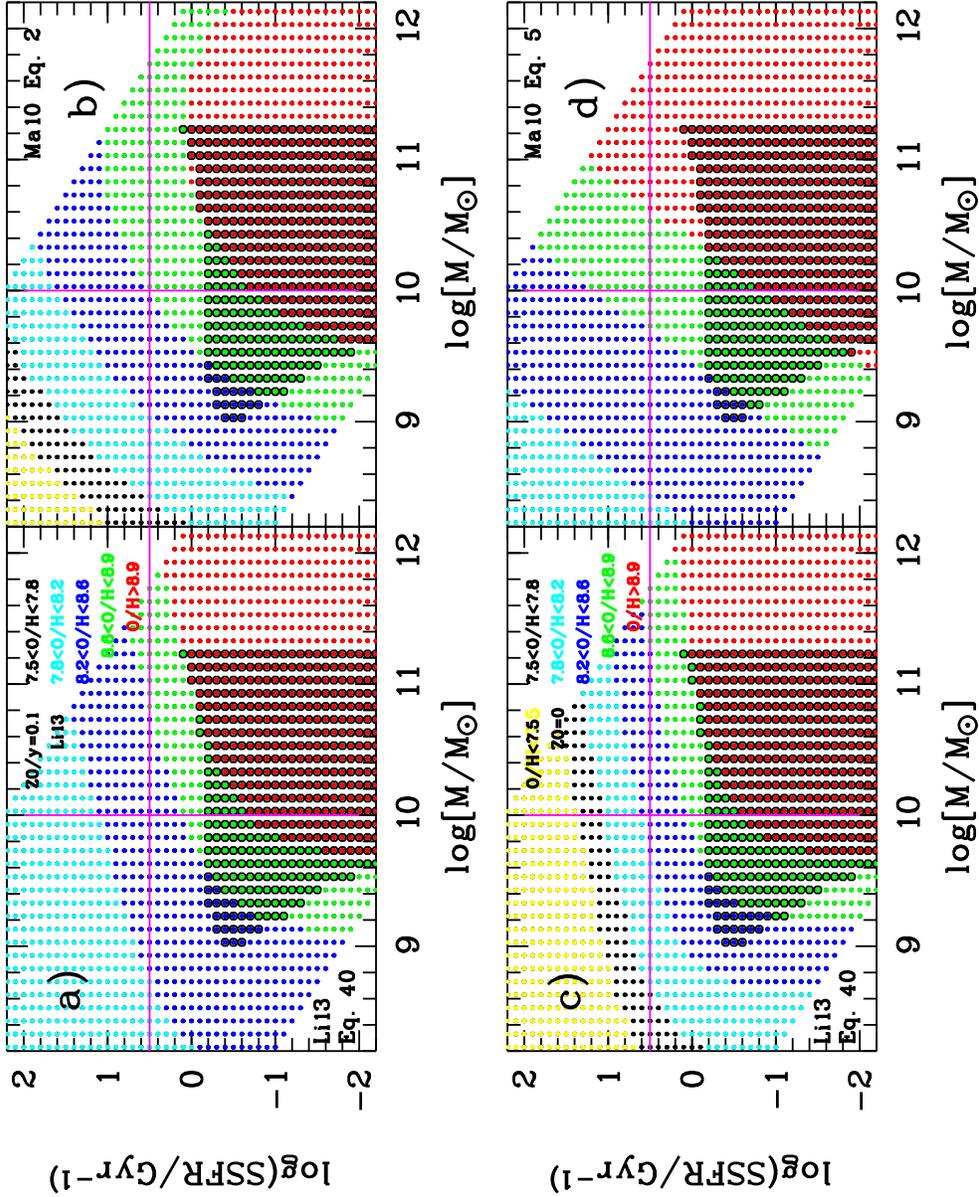}
\caption
{
\label{FMRMa10Li13} 
\footnotesize 
The SSFR - mass relation color-coded according to the expected metallicities derived using Eq.\,40 of Li13, with a redshift independent $\epsilon(M)$ and the parameters from Tab.\,1 in Li13 for different infall metallicities $Z_{0}$ relative to the yield $y$, $Z_{0}/y=0.1$ (panel a) and $Z_{0}/y=0$ (panel c), and using Eq.\,2 (panel b) and Eq.\,5 (panel d) from Ma10.
The black open circles in all panels show the parameter space in SFR and mass for which the predicted O/Hs from Li13 and Ma10 differ by less than 0.05\,dex.
This roughly corresponds to the range in observed SFR 
and mass values of SDSS galaxies, from Tab.\,1 in Ma10, which were both used to fit the parameters in Eq.\,40 of Li13 and to derive Eqs.\,2 and 5 of Ma10.
}
\end{figure}

2. The mean MZR at $2.1<z<2.5$ is shifted to lower metallicities by a factor of three to five compared to  the local SDSS relation (Fig.\,\ref{MassOH}).
This shift is larger than the  factor of two evolution found by Eb06.
A possible cause for this is the unknown type-2 AGN contamination for 83 out of 87 galaxies in the Eb06 sample, with too high \NII\, emission lines possibly mimicking too high metallicities.
Another cause is the strong ionization parameter dependence of the \citet{petpag04} method, especially at higher metallicities like the ones in the Eb06 sample (see Fig.\,\ref{N2_IonPar}), which can produce large errors in the metallicity measurements.

\begin{figure}[h]
\includegraphics[width=14cm,angle=0,clip=true]{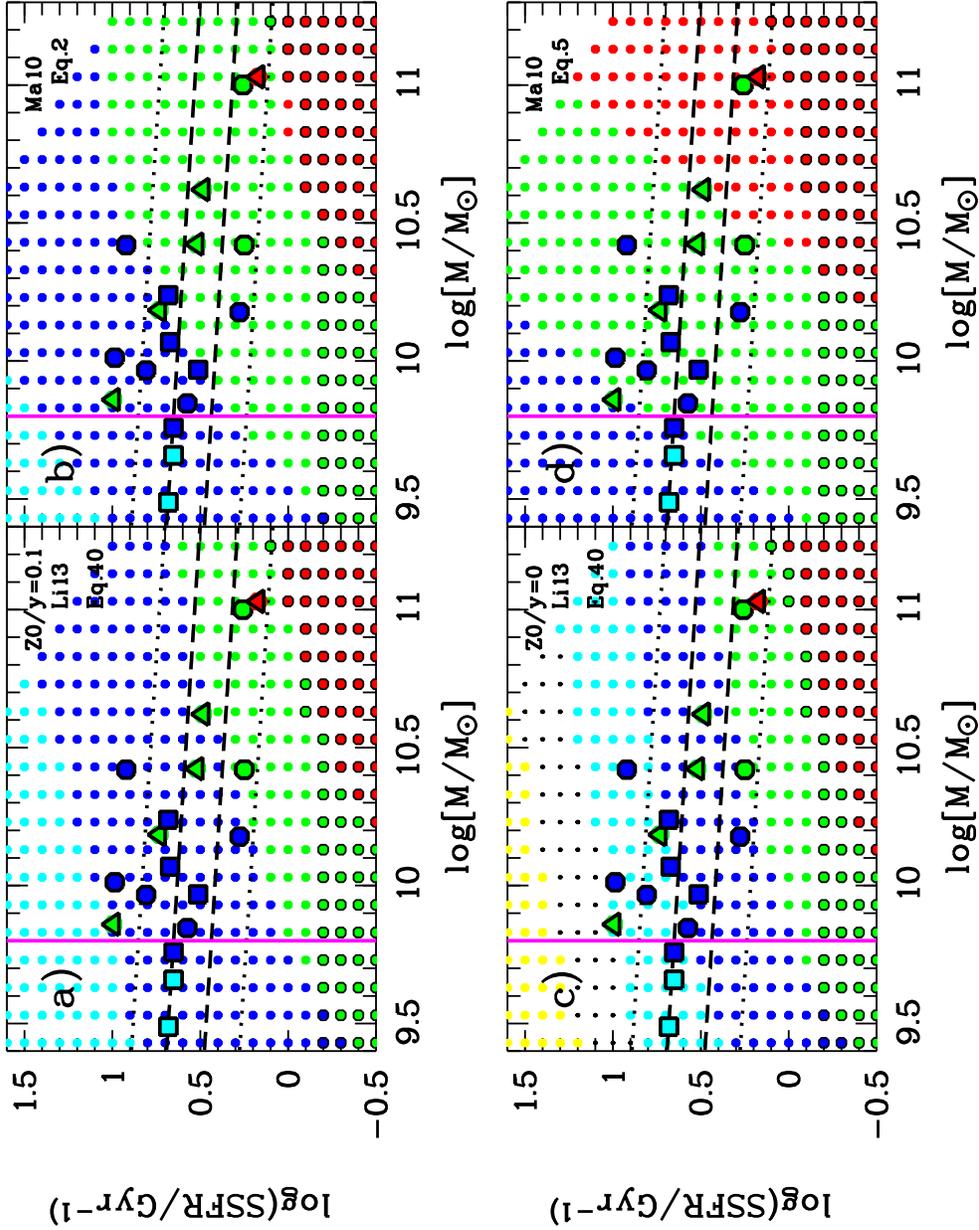}
\caption
{
\label{SFRMass_Obs} 
\footnotesize 
The colors of the grid points indicate, like in Fig.\ref{FMRMa10Li13}, the metallicities expected for the different Z(M,SFR) extrapolations of Ma10 and Li13. 
Additionally,  galaxies with NIR spectroscopy at $z>2$ are overplotted: zCOSMOS galaxies (filled circles), Eb06 data (filled triangles), and the \citet{cullen13} stacked measurements (filled squares). 
All data points are color-coded according to their \emph{measured} metallicities (e.g., green color for $8.6<O/H<8.9$, see legend in Fig.\,\ref{FMRMa10Li13}), to explore if the measured metallicities (color of the \emph{data} points) and the expected metallicity assuming a redshift independent Z(M,SFR) relation (color of the \emph{grid} points) are in agreement.
}
\end{figure}

3. We find direct indications that SFR is still a second parameter in the MZR at $z>2$, in the sense that galaxies at $2.1<z<2.5$ with higher metallicities have lower (S)SFRs (see Fig.\,\ref{SFRMass_Obs}). 

4. The metallicities expected at high redshift for a non-evolving FMR up to $z\sim 2.5$ depend on the formulation and thus the extrapolation of the low redshift Z(M,SFR) relation, since the high redshift galaxies lie in a different part of the (M,SFR) plane.  The metallicities reported here and in \citet{cullen13} are more consistent with a non-evolving FMR if the physically-motivated  Z(M,SFR) from Li13 is used for the extrapolation rather than the empirical Z(M,SFR) relation from Ma10.

5. This study emphasizes the difficulties of measuring metallicities using three to five emission lines at $z>2$: we could obtain non-degenerate oxygen abundances for only 7 out of 20 galaxies (35\% of the sample) with the achieved quality of the data. The error bars of some emission lines, especially [NII] and \Hb, are quite large, resulting in quite large error bars for the oxygen abundances. This indicates that longer exposure times and/or more sensitive detectors are necessary to better constrain the metallicities of high-redshift galaxies. Therefore, much larger samples with NIR spectroscopy of galaxies at $z>2$, which now become feasible with the advance of multi-object spectrographs like  FMOS on Subaru, MOSFIRE on Keck, or KMOS on VLT, are needed to put these findings on solid footings.

\acknowledgments
We are grateful the anonymous referee for his/her suggestions which have improved the paper. We greatly acknowledge the contributions of the entire COSMOS collaboration, consisting of more than 70 scientists.

%
%

%
%

\clearpage
\begin{deluxetable}{cccccccccccc}
\tabletypesize{\footnotesize}
\setlength{\tabcolsep}{0.05in}
\tablewidth{0pt}
\rotate
\tablecaption{\label{SINF_MOIRCS} \footnotesize Observed and derived quantities for the 
 zCOSMOS $z > 2$ galaxies
  }
\tablehead{
\colhead{Id}     &    \colhead{$z_{UVabs}$}   &
\colhead{$z_{EL}$} &  \colhead{[OII]\tablenotemark{a}}    &   \colhead{\Hb\tablenotemark{a}}
  &  \colhead{[OIII]\tablenotemark{a}}    &  \colhead{\Ha\tablenotemark{a}}    & \colhead{[NII]\tablenotemark{a,b}}     &
  \colhead{O/H}
  &     \colhead{A$_{\rm V}$}&
  \colhead{Instrument}&
  \colhead{log(M/M$_{\sun}$)}
}
\startdata
 401925&2.1394& 2.1418&	OH                &  7    $\pm$  2     &  42   $\pm$ 3   &   20  $\pm$   2   &   $<$1             &8.44$^{+0.14}_{-0.31}$&0.03$^{+0.23}_{-0.03}$&SINFONI&9.97\\	
 404360&2.2451& 2.2453&	No J spectrum    &   No H spectrum  &     No H spectrum&   12 $\pm$   1 &   $<$1&-&-&SINFONI&10.23\\
 404724&2.4750& 2.4771&	20   $\pm$   3    &  8\tablenotemark{c}  $\pm$  3  &  33  $\pm$ 2     &   30  $\pm$   3   &OH&8.64$^{+0.16}_{-0.23}$&0.49$^{+0.30}_{-0.49}$&SINFONI&11.00\\    
 407711\tablenotemark{e}&2.1062& 2.1080&	$<$5                &  5.3\tablenotemark{c} $\pm$1.7    &  22\tablenotemark{d}    $\pm$ 3 &   20 $\pm$   2 &   $<$1.7&8.70$^{+0.08}_{-0.93}$&0.85$^{+1.05}_{-0.77}$&SINFONI&10.63\\
 408147\tablenotemark{e}&2.1718& 2.1746&	14    $\pm$  1    &11.5\tablenotemark{c} $\pm$  4  &  14  $\pm$ 3     &   45  $\pm$   3&   6$\pm$0.5 &8.53$^{+0.49}_{-0.04}$&2.50$^{+0.15}_{-2.50}$&SINFONI&10.78\\
 408530\tablenotemark{e}&  2.2841&  2.2834&  4.2   $\pm$  0.6 &   3.7\tablenotemark{c}  $\pm$  2    &     21\tablenotemark{d}   $\pm$ 4    &    22 $\pm$    4  & $<$2.9 &8.60$^{+0.17}_{-0.52}$&1.75$^{+0.55}_{-1.01}$&MOIRCS&10.29\\
 408609&  2.4425&  2.4466&  1.5 $\pm$    0.5 &    OH   &     OH    &    7.8$\pm$    1&  OH &-&-&MOIRCS&10.17\\
 408888&  2.3757&  2.3783&  3.3 $\pm$    0.5 &    OH             &    OH   &    13.8 $\pm$  1.7  &  OH &-&-&MOIRCS&10.47\\
 409130\tablenotemark{e}&  2.2784&  2.2837&  2.8   $\pm$  0.3 &    1.5  $\pm$  	0.5 &	  8   $\pm$   2   &    6  $\pm$    1  & $<$0.5&8.49$^{+0.24}_{-0.34}$&0.61$^{+0.46}_{-0.61}$&MOIRCS&10.40\\
 409585&2.1909& 2.1943&	$<$4.2              &  3.0  $\pm$  1.0   &  20.5   $\pm$ 1   &   9   $\pm$   1.0 &   $<$0.8&8.40$^{+0.13}_{-0.26}$&0.02$^{+0.28}_{-0.02}$&SINFONI&10.18\\ 
 409985&2.4505& 2.4579&	12   $\pm$   2    &  3.5    $\pm$  1.0   &  22   $\pm$ 2   &   11.2$\pm$	0.8  &   1  $\pm$   0.2&8.59$^{+0.05}_{-0.06}$&0.05$^{+0.05}_{-0.05}$&SINFONI&10.42\\
 410123&2.1993& 2.1993&	10  $\pm$   2     &  3.6\tablenotemark{c}    $\pm$  1.3     &  7 $\pm$  2     &  14 $\pm$   2     &   OH&8.45$^{+0.16}_{-0.23}$&1.13$^{+0.38}_{-0.46}$&SINFONI&10.01\\
 410432&2.4498& 2.4556&	No J spectrum    &   No H spectrum  &     No H spectrum&   7 $\pm$   1 &   $<$0.9&-&-&SINFONI&10.81\\
 411785&2.2429& 2.2450&	17   $\pm$   3    &  18   $\pm$  2     &  110  $\pm$ 10  &   55  $\pm$	 5   &	  4  $\pm$   1.0&8.57$^{+0.06}_{-0.11}$&0.25$^{+0.32}_{-0.25}$&SINFONI&10.42\\
 412952\tablenotemark{e}&2.3183& 2.3211&	No J spectrum    &  2.7  $\pm$  1.0   &  7  $\pm$ 0.5 &  12  $\pm$ 4    &   OH&8.50$^{+0.41}_{-0.23}$&1.41$^{+1.14}_{-1.41}$&SINFONI&10.44\\
 428671&  2.2969&  2.2953&  OH               &  No H spectrum & No H spectrum       &    2.8$\pm$    0.9&  OH &-&-&MOIRCS& 9.80\\
 429111\tablenotemark{e}&  2.2750&  2.2816&  OH               &    1.2   $\pm$ 0.5  &     3   $\pm$   0.5 &    4.8$\pm$    1&  $<$0.6&8.35$^{+0.56}_{-0.53}$&1.13$^{+1.15}_{-1.13}$&MOIRCS& 9.74\\
 429152&  2.2806&  2.2811&  6   $\pm$    2   &    2.1  $\pm$   0.2 &	  13.5  $\pm$   0.5   &    7  $\pm$   0.3 & $<$0.3&8.42$^{+0.09}_{-0.13}$&0.11$^{+0.25}_{-0.11}$&MOIRCS& 9.85\\
 429182&  2.3750&  2.3801&  Bad pixels area  &     1.4\tablenotemark{c}  $\pm$  0.5                &     5.1 $\pm$   0.5 &    5.5$\pm$    1.3& OH &-&-&MOIRCS& 9.97\\
 429868\tablenotemark{e}&2.4414& 2.4469&	No J spectrum   &  2.5\tablenotemark{c}  $\pm$  1   &  9  $\pm$2 &   10 $\pm$   2 &   $<$1.7&8.58$^{+0.29}_{-0.61}$&1.00$^{+1.01}_{-1.00}$&SINFONI&10.07\\
\enddata
\tablenotetext{a}{Fluxes are given in $10^{-17}\rm{ergs}\,\rm{s}^{-1}\rm{cm}^{-2}$}
\tablenotetext{b}{$3\sigma$
  upper limits for the emission line flux of \NII\, are given if \NII\, is not detected}
\tablenotetext{c}{\Hb\, of these objects were too faint to be measured, or coincide with
a strong OH skyline, so the respective \Hb\, flux could not be
measured. We calculated  \Hb\, fluxes for these galaxies assuming
$0<A_{V}<2$ and case B recombination and use these to derive O/Hs, as discussed in Sect.\,\ref{emline}, and these \Hb\, flux values are given in the table.
}
\tablenotetext{d}{This \OIIIa\, flux was derived by multiplying the \OIIIb\, flux by a factor of three, because \OIIIa\, is
  affected by a strong OH skyline, but \OIIIb\,  is not.}
\tablenotetext{e}{Double-valued O/H solution.} 
\end{deluxetable}

%
%

\clearpage

%
%

\clearpage

\clearpage


\clearpage

\clearpage


\clearpage


\end{document}